\documentclass[smallextended,table]{svjour3}
\usepackage[utf8]{inputenc}
\usepackage{todonotes}
\usepackage{breakcites}
\usepackage{natbib}
\usepackage{makeidx}  

\usepackage[hyphens]{url}
\usepackage{longtable}
\usepackage{csquotes}
\usepackage{graphicx}
\usepackage{ulem}
\usepackage{tikz}
\usetikzlibrary{mindmap,trees}
\usetikzlibrary{calc,positioning}
\definecolor{rosso}{RGB}{220,57,18}
\definecolor{giallo}{RGB}{255,153,0}
\definecolor{blu}{RGB}{102,140,217}
\definecolor{blu2}{RGB}{0, 137, 204}
\definecolor{blu3}{RGB}{17, 102, 139}
\definecolor{verde}{RGB}{16,150,24}
\definecolor{verde2}{RGB}{117, 200, 16}
\definecolor{viola}{RGB}{153,0,153}
\definecolor{viola2}{RGB}{83,40,213}
\definecolor{myellow}{RGB}{255,233,114}

\usepackage{rotating}
\usepackage{lscape}
\usepackage{caption}
\usepackage{multirow}
\usepackage{booktabs}
\usetikzlibrary{arrows.meta}
\providecommand{\JEL} [1] {\textbf{JEL Classification} #1}

\usepackage{etoolbox}
\newcommand*{\affaddr}[1]{#1} 
\newcommand*{\affmark}[1][*]{\textsuperscript{#1}}

\makeatletter
\newcommand{\printfnsymbol}[1]{%
  \textsuperscript{\@fnsymbol{#1}}%
}
\makeatother

\def\makeheadbox{{%
\hbox to0pt{\vbox{\baselineskip=10dd\hrule\hbox
to\hsize{\vrule\kern3pt\vbox{\kern3pt
\hbox{\bfseries [Digital Finance]}
\hbox{This is a post-peer-review, pre-copyedit version of this article.}
\hbox{The final authenticated version is available online at: https://doi.org/10.1007/s42521-021-00035-5}
\kern3pt}\hfil\kern3pt\vrule}\hrule}%
\hss}}}

\begin{document}

%
%
\title{A Blockchain-based Forensic Model for Financial Crime Investigation: The Embezzlement Scenario }
\titlerunning{Blockchain Forensics for Embezzlement Investigation}  
%

\author{%
Lamprini Zarpala\protect\affmark[1] \and Fran Casino\affmark[2]\textsuperscript{,}\affmark[3] }
\authorrunning{Lamprini Zarpala  \and Fran Casino}

\institute{
              Lamprini Zarpala
              \email{zarpala@unipi.gr}           
           \and
           Fran Casino
              \email{francasino@unipi.gr} \\
              \affaddr{\affmark[1] Department of Banking and Financial Management, Piraeus University, Piraeus (Greece)}\\
\affaddr{\affmark[2] Department of Informatics, Piraeus University, Piraeus (Greece)}\\
\affaddr{\affmark[3] Athena Research Center, Athens (Greece)} \\
\and
*All authors equally contributed to this work
}

\date{Received: 23 August 2020 / Accepted: 30 June 2021}




\maketitle            

\begin{abstract}
The financial crime landscape is evolving along with the digitisation of financial services. Laws, regulations and forensic methodologies cannot efficiently cope with the growth pace of novel technologies, which translates into late adoption of measures and legal voids, providing a fruitful landscape for malicious actors. In this regard, the features offered by blockchain technology, such as immutability, verifiability, and authentication, enhance the robustness of financial forensics.
This paper provides a taxonomy of the prevalent financial investigation techniques and a thorough state-of-the-art of blockchain-based digital forensic approaches. Moreover, we design and implement a forensic investigation framework based on standardised procedures and document the corresponding methodology for embezzlement scheme investigations. The feasibility and adaptability of our approach can be extended and embrace all types of fraud investigations and regular internal audits. We provide a functional Ethereum-based implementation, and we integrate standardised forensic flows and chain of custody preservation mechanisms. Finally, we discuss the challenges of the symbiotic relationship between blockchain and financial investigations, along with the managerial implication and future research directions.   

\JEL {K42 $\cdot$ K41 $\cdot$ M42 $\cdot$ G21}
\end{abstract}

\section{Introduction}

 The impact of fraud and economic crime on all organizations worldwide still reaches high-record levels. The most frequently committed fraud scheme is asset misappropriation, where an employee is stealing or misusing organizational resources \citep{ACFE2020}. A provocative question arising when it comes to fighting fraud is about the deployed technologies during the investigation.
 
Data acquisition is one of the most critical steps during forensic investigations. This is interpreted as maintaining a chain of custody for data and performing data integrity validation to ensure tamper-proofness \citep{1393,Gottschalk2018}.
Advanced analytic techniques, such as machine learning, cognitive computing, and, in general, automated data processing methods, are some of the trends in forensics\footnote{Financial forensics is a methodology which combines investigation and auditing skills to identify evidence for a potentially fraudulent activity that might end up in litigation. Fraudulent activities might derive either from within or outside the organization.} in the financial services sector.
 
In this work, we choose a real-world embezzlement\footnote{It is the misappropriation of funds that have been entrusted to an employee for care or management} scenario and apply the forensic-by-design architecture of blockchain to validate the audit trail's integrity. To uncover embezzlement schemes, investigators have developed a hypothesis about the sequence of events and applied specific fraud analytic tools presented in Section (\ref{Invest_tech}). The results from the data forensics analysis and the other sources of documentation such as hard copies and CCTV tapes are stored in the audit trail, supporting the hypothesis in potential litigation proceedings \citep{Examiners2020,Guide2012,1393}.

The sequence of custody, control, transfer, and disposition of the audit trail of evidence is called \textit{chain of custody.} The chain of custody indicates who had access to records and thus the degree of difficulty in manipulation. It assures that evidence is not damaged or altered in any way since gaps in the chain may illustrate mishandling of evidence which leads to damaging the case \citep{Jwbt2012}. The importance of chain of custody is highlighted when the fraud investigator has to perform one of the most challenging tasks to put financial information in simple terms to be understandable from decision-makers (senior management, prosecutor, etc.). The collection, examination, analysis, interpretation, and presentation of digital evidence shapes investigations are crucial in proving embezzlement cases where digital evidence is prevalent because the employee under investigation may be well aware of the bank's systems, procedures, and control gaps. In this regard, recent work in the digital forensic community has established reliable scientific methodologies and common standards in its workflows \citep{biasiotti2018handling,goblerstandards}. However, it still faces many challenges due to the volatile and malleable nature of the evidence and the continuous advances in technology that introduce new attack vectors \citep{huang2018systematically,casino2019immutability}. 


In parallel to the digitization of financial services, blockchain technology \citep{nakamoto2008bitcoin} has experienced a wide adoption due to the myriad of possibilities and applications that it enables \citep{casino2019systematic,crosby2016blockchain}. Setting aside the different categorizations of blockchain in terms of applications and consensus algorithms (i.e. the methodology to reach an agreement between the different participants of the blockchain network to validate a transaction), they offer different features, such as auditability, security, decentralization and transparency \citep{xu2019systematic,casino2019systematic}. In addition, blockchains also provide immutability \citep{politoublock,casino2019immutability}, which is an interesting feature, yet poses significant challenges to principles such as the right to be forgotten or the EU General Data Protection Regulation Directive (GDPR) \citep{politou2018forgetting}. 

Blockchain applications in the financial context are already a fact \citep{guo2016blockchain,peters2016understanding,casino2019systematic}. Nevertheless, blockchain capabilities can also be used maliciously to enable cryptocurrency fraud scams such as investment and ponzi schemes, embezzlement, phishing, and ransomware \citep{cryptostealing2020}. Therefore, investing efforts to prosecute fraud as well as its prevention in both traditional and next generation systems is mandatory. In this regard, the literature is still scarce in blockchain-based methods oriented to fraud prevention \citep{aicpas2019fraud,8883647}. Promising research lines such as crypto de-anonymization methods have the potential to enable successful investigations and prosecutions \citep{shentu2015research}, yet such solutions have many constraints, as well as not being feasible in cryptocurrencies such as Monero. 

Based on the above requirements, it is apparent that the blockchain features may enhance forensic procedures \citep{Al-Khateeb2019149}. For instance, the immutability of blockchain can guarantee the verifiability of the chain of custody of evidences and provide an auditable trail of events. The latter is mandatory to be presented as a solid proof in a courtroom.

\subsection{Motivation and Contribution}

The amount of cyber challenges faced by the financial sector requires cooperation between different entities and actors both at national and international levels. In this regard, several cybersecurity initiatives in the finance sector are being pushed by organizations such as ENISA \citep{enisainitiatives21} to establish a standard and fertile ground in such a critical sector. Therefore, the work presented in this paper contributes to reinforcing key dimensions such as information sharing and capacity building, awareness and training, standardization and certification, and research and innovation.

In this paper, we address the relevance of blockchain technology and its inherent features for the management of evidence in forensic investigations. We are concerned with a case study that occurs within a financial institution. This type of fraud is also known as bank fraud. According to the Basel Committee on Banking Supervision, there are different kinds of fraud: internal/occupational frauds or external frauds. In this context, we will examine occupation fraud as an embezzlement scheme for unauthorised withdrawals. The detection method used is a review of the source documentation that might confirm or refute the allegation of embezzlement.  
In addition, we provide a taxonomy of the financial investigation techniques, and comprehensive overview of the state of the art in blockchain forensics, and we propose a blockchain-based architecture for embezzlement' fraudulent activities enabling sound preservation of the chain of custody. This architecture and the forensic flows are implemented according to well-known standards and guidelines, such as the ones described in Section \ref{sec:forensic}. Moreover, we provide an implementation based on Ethereum and smart contracts to preserve the chain of custody as well as the trail of events. Overall, our solution enables various features and benefits, such as integrity verification, tamper-proof, and future enhancement of similar investigations. Therefore, approaches such as the one proposed in this work aim to close the gap between the technicalities of financial investigations and legal procedures, enhancing the robustness of the current state of practice. In addition, we discuss several measures to improve the adaptability of our system towards similar crimes, and we provide a fertile ground for further research.

To the best of our knowledge, this is the first work that provides a blockchain-based forensic sound procedure for embezzlement and an implementation based on smart contracts. Even though we use the embezzlement scenario to illustrate the forensic procedure, our high-level architecture can be extended in all fraud investigations and also in regular internal audits.

\section{Related works}

\subsection{Financial Investigation Techniques} \label{Invest_tech}
\begin{figure}[th]
    \centering
    \includegraphics[width=\textwidth]{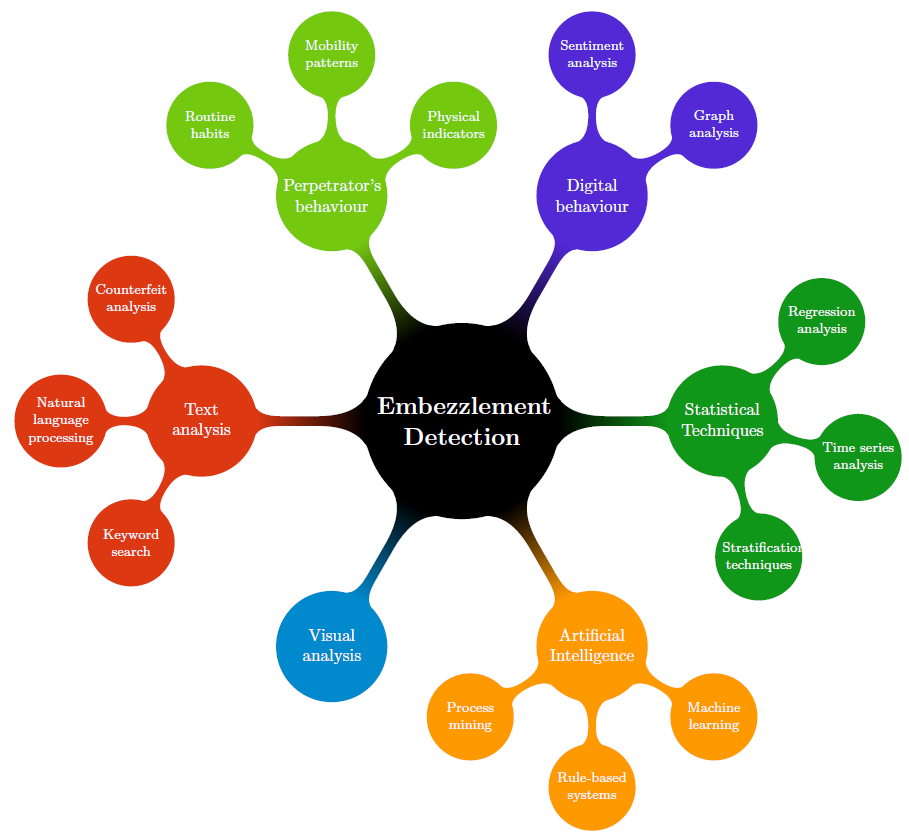}
    \caption{Mindmap abstraction of the different families of tools that enable financial investigation.}
    \label{fig:financetikz}
\end{figure}

We analysed the state of the art of financial investigation techniques and tools and leveraged a taxonomy of the different families of methods that can be used according to the elements and actors under investigation. Figure \ref{fig:financetikz} provides an overview of the taxonomy and the most relevant techniques related to the effective design of fraud detection.

 Fraud-related technical literature incorporates 
in the design of detective controls the importance of red flags for perpetrator's behaviour, which can validate the predictive theory \citep{GULLKVIST201344}. In the embezzlement case, and in asset misappropriation in general, the detective controls rotate around employee's financial status, consumption habits, egocentric behaviour analysis, performance evaluation, adherence to vacation policies, job dissatisfaction \citep{SINGLETON2010}.

To detect deviations and specific behaviour in documents and social networks \citep{yoo2009comparison}, fraud investigators use techniques of sentiment analysis \citep{Goel2016} augmented with graph analysis \citep{pourhabibi2020fraud} to discover the degree of connectivity between individuals. Graph analysis is also used to evaluate the performance of a small subset of transactions and to identify the accounts involved in a cross-channel transaction. The latter cannot be identified with classical statistical source-account profiling \citep{10.1007/978-3-662-54970-4_2}.
 
 The visual analysis category mainly refers to the graphical representation of numerical and categorical data, usually in form of diagrams or plots (e.g. bar, radial, box plots, heat maps), and 3D visualizations. This is a well-known approach for fraud detection used by investigators to speed up manual analysis of data since this approach enhances its readability \citep{LEITE2018198}. An essential aspect of visual analysis is the interaction of such systems, letting investigators select, explore, and filter the visualized data.

Statistical and data analytics are leveraged with an extensive set of techniques \citep{kobayashi2011probability,BANARESCU20151827} which aim to detect anomalies in data. These anomalies could indicate customers' abnormal behavior (e.g frequency and amount of regular transactions,
changes in geography, and outlier detection) \citep{cliff2017statistical,Ngai2011}.

Artificial intelligence is widely used in the finance field to detect fraud schemes such as embezzlement \citep{choi2018artificial,awoyemi2017credit}. One of the main approaches is leveraging pattern recognition by using machine learning techniques (both supervised and unsupervised) to classify data according to some criteria (e.g., identify multiple transactions at a customers' account in quick succession) \citep{sadgali2019performance} and to predict the occurrence of fraud through the use of neural networks, Bayesian learning, decision trees and association rules \citep{CERULLO199914,CERULLO199914a}. Similarly, clustering algorithms are used to explore how data are related and find further insights. Other approaches also include rule-based expert systems and process mining models which can extract and detect further data patterns \citep{omair2020systematic}.

Finally, in parallel to the aforementioned techniques, text analysis is another relevant source of evidence. Due to the modus operandi of the malicious actors, the use of natural language processing to detect authorship of documents \citep{de2001mining,green2013comparing}, including the use of specific keywords to tag specific transactions or operations, can help in the detection of embezzlement \citep{markowitz2014linguistic}. Moreover, automated systems to detect forged documents and signatures \citep{gideon2018handwritten} are critical in this context.

For our analysis, we reviewed prior researches related to embezzlement schemes from the finance-related literature. Nevertheless, the existing literature gives little attention to the embezzlement schemes performed within financial institutions.
Embezzlement schemes are more likely to be realised in situations when either the separation of duties or the relevant audit trails are weak or nonexistent. One of the most common types of banking fraud is the unauthorized withdrawals from customer accounts \citep{Examiners2020} and may take various forms. For instance, it could be stealing money from customers by providing encoded deposit slips \citep{Manning2011} or borrowing from today's account's receivable and replace them with tomorrow’s receipts. In all cases, a scheme to be successful requires the creation of false data, reports, or data entries \citep{Bologna1995}. The aforementioned tools and techniques can be applied to detect embezzlement schemes. Yet, one of the main challenges in the financial sector is accommodating such solutions to each organization's systems. Moreover, as stated in Section 1.1, there are several challenges in terms of interoperability, training, policies and regulations, and standardization, that prevent the creation of common frameworks to exploit the full potential of such techniques.

\subsection{Blockchain Forensics}
Despite the suitability of blockchain towards preserving a trail of events in an immutable and verifiable manner, only recently authors have started to explore it. 

\begin{figure}[th]
    \centering
    \includegraphics[width=\columnwidth]{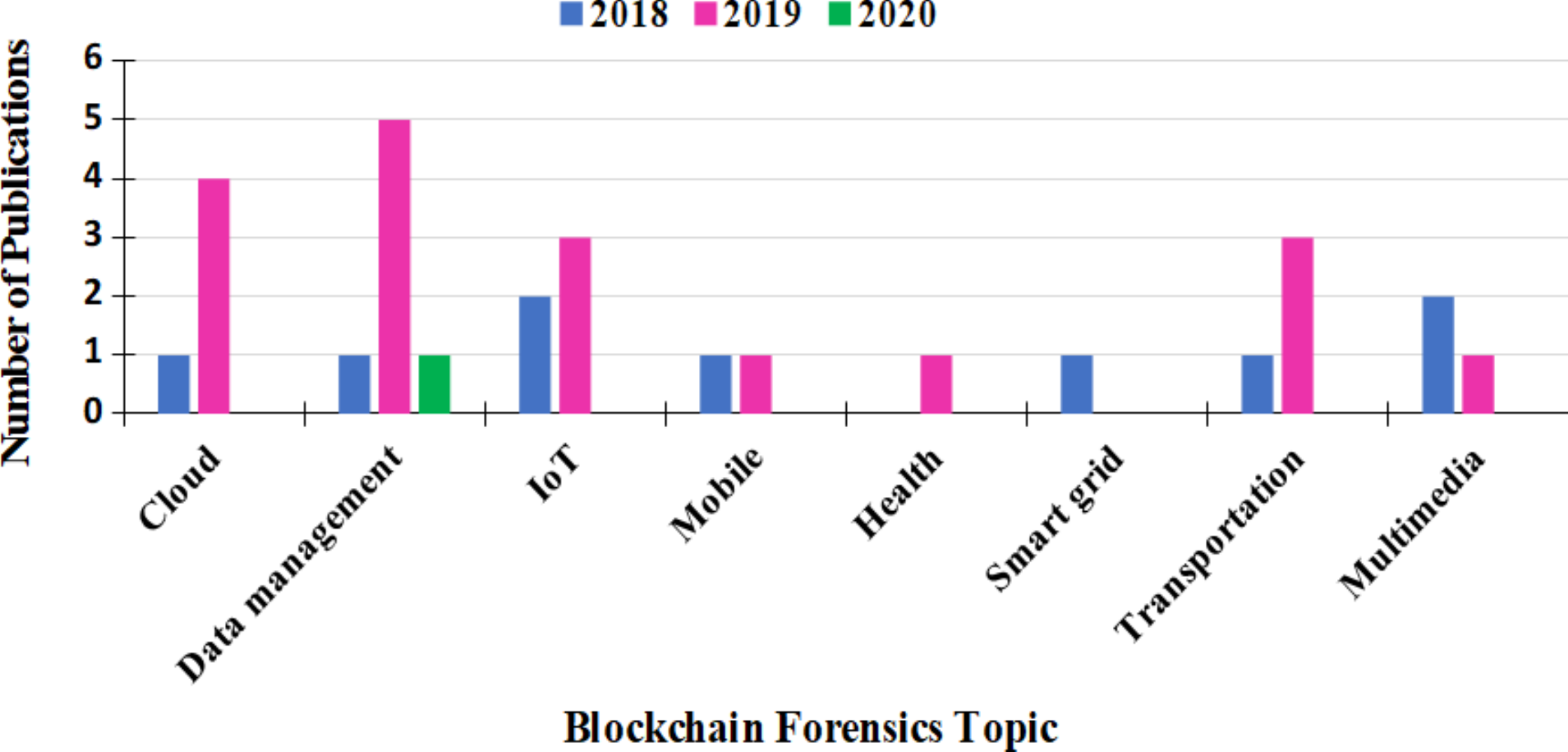}
    \caption{Summary of blockchain forensic topics and the number of publications per year.}
    \label{fig:soa}
\end{figure}

In order to retrieve the relevant literature in blockchain, we queried the Scopus database by using the keywords \enquote{blockchain} and \enquote{forensics}, and we located further studies by means of the snowball effect (i.e. additional literature referenced by the articles found in the initial search). In total, 28 articles were selected according to a specific criteria (i.e. we peer-reviewed and excluded some papers based on their structural quality, language, and subject area). Next, we performed a keyword-based and topic classification, depicted in Figure \ref{fig:soa}. Therefore, we identified eight different blockchain forensic application topics which span from 2018 to June 2020, being the most populated the ones related with data management, IoT, and cloud. 

In the context of cloud forensics, a framework to enable fast incident response was proposed in \cite{Ricci2019}. In \cite{Zhang2018}, the authors created a collaboration and evidence management framework to improve the coordination of investigations when different stakeholders are involved. In the case of \cite{Rane2019} and \cite{Duy2019416}, authors focused in logging-as-a-service tools for securely storing and processing logs while coping with issues of multi-stakeholder collusion, and the integrity and confidentiality of logs. In this context, security and robust access mechanisms are mandatory, as discussed in \cite{Pourvahab2019153349}.

The works classified in data management topic present different models for data processing based on scalable solutions such as permissioned blockchains \citep{Gopalan2019,Lone2019,Xiong2019} and blockchains using lightweight consensus mechanisms \citep{Tian2019}. Other relevant features showcased by authors are the verifiability of the trail of events \citep{Weilbach2019,Bonomi2020} and the classification of the evidence in terms of features, enabling further data processing \citep{Billard2018}.

The evidence collected from IoT devices and the interactions between different stakeholders are one of the main features studied in the literature \citep{Brotsis2019,Hossain2018,Ryu2019,Li20191433}. 
In addition, privacy-preserving identity management \citep{Le2019} is a mandatory feature to be considered in such context.

The blockchain-based mobile forensic research focuses on applications and malware detection. In this regard, authors propose the use of consortium blockchains to detect malware and based on statistical analysis of each application's feature \citep{Gu2018,Homayoun2019}.

The multimedia forensic topic includes works related with processing and storing CCTV video evidence \citep{Kerr2019}, the multimedia evidence captured by smartphones \citep{Samanta2018} and image-based provenance and integrity \citep{Zou2019}.

In the healthcare context, the work presented in \cite{malamas2019forensics} proposes a blockchain-enabled authorization framework for managing both the Internet of Medical Things (IoMT) devices and healthcare stakeholders. 
Smart grid forensics and its relevance is discussed in \cite{Kotsiuba2019}. Moreover, the authors showcase the benefits of blockchain towards enhancing energy optimization, security and managerial tasks.

Transportation forensics has captured the interest of researchers due to its timely relevance due to its seamless relationship with emerging technologies such as self-driving automation, IoT-based sensing, and 5G communication networks \citep{Patsakis2019}. Nevertheless, such novel frameworks and the adoption of new data privacy frameworks (like the GDPR) call for the development of sound forensic mechanisms to analyze traffic accidents and protect users' sensitive data. For instance, in \cite{Billard2019}, the author proposes a system to manage user's requests and their compliance with legal frameworks.

A privacy-preserving framework is also proposed in \cite{Kevin2019}, this time for managing sensitive navigation data while ensuring user's anonymity. In \cite{Cebe2018}, the authors propose a blockchain-based forensics system that enables traceable and privacy-aware post-accident analysis with minimal requirements in storage and processing. In \cite{Patsakis2019} a blockchain-based framework is proposed for keeping logs of all hardware profile changes and updates in a vehicle.

Finally, we analysed the maturity of the literature according to their development stage in Table \ref{tab:maturity}. In general, we may observe that the majority of solutions are in an early stage and thus, more efforts need to be devoted to this research field to exploit all the potential of blockchain. A more detailed exploration of the blockchain forensics literature can be found in \cite{dasaklissokforensics}

\begin{table}[ht]
\caption{Maturity of blockchain-based forensics literature.}
\resizebox{\textwidth}{!}{
\begin{tabular}{c|c|c}
\toprule   
\textbf{Implementation \& Tests} & \textbf{Partial Implementation} &\textbf{Not Provided} \\
\midrule 
\cite{Rane2019}     &                \cite{Li20191433}           &  \cite{Ricci2019}\cite{Gopalan2019}\\

\protect\cite{Duy2019416}  &  \cite{Gu2018}  &  \cite{Billard2018} \cite{Hossain2018} \\

\protect\cite{Tian2019} \cite{Bonomi2020} &   \cite{Ryu2019}  &  \cite{Brotsis2019}  \cite{Billard2019} \\

 \cite{Zhang2018}  \cite{Patsakis2019}  &    &\cite{Kevin2019} \\

 \cite{Samanta2018} &   &  \cite{Kotsiuba2019}  \cite{Cebe2018} \\

  \cite{Pourvahab2019153349}  &   &  \cite{Weilbach2019}   \\

  \cite{malamas2019forensics}  \cite{Zou2019}  & &  \cite{Xiong2019} \\

\cite{Kerr2019} \cite{Lone2019}  &   &  \cite{Le2019} \cite{Homayoun2019}   \\
\bottomrule
\end{tabular}
}
  \label{tab:maturity}
\end{table}

\subsection{Forensic Models}
Despite that one of the main issues in the digital forensics research field is the standardisation of procedures according to each topic or context \citep{GARFINKEL2010S64}, there exist several well-known forensic guidelines and models \citep{forensic_guidelines,kent2006sp,interpolguidelines19}, which are summarised in Table \ref{tab:mainmodels}. Moreover, a review of the international development of forensic standards, such as ISO/IEC 27043:2015, can be found in \cite{WILSONWILDE20181} and \cite{boasiako17}.

\begin{table}[ht]
   \setlength{\tabcolsep}{7pt}
   \scriptsize
   \caption{Most well-known forensic models and guidelines.}
   \resizebox{\textwidth}{!}{
  \begin{tabular}{p{.75\textwidth}cc}
     \toprule
   \textbf{Name} &  \textbf{Year} &  \textbf{Reference} \\
   \midrule
Digital Forensic Investigation Model  &  2001  & \cite{kruse2001computer}\\
Digital Forensic Research Workshop & 2001 &\cite{palmer2001road}\\
Abstract Digital Forensic Model & 2002 &\cite{reith2002examination}\\
Integrated Digital Investigation Model & 2004 &\cite{carrier2003getting}\\
Enhanced Digital Investigation Process Model& 2004 &\cite{baryamureeba2004enhanced}\\
Extended Model of Cybercrime Investigation & 2004 & \cite{ciardhuain2004extended} \\
NIST Guide to Integrating Forensic Techniques into Incident Response & 2006 &\cite{kent2006sp}\\
Digital Forensic Model for Digital Forensic Investigation &2011 &\cite{ademu2011new}\\
International Organization for Standardization ISO/IEC 27043:2015 & 2015 & \cite{iso270432015} \\
INTERPOL Guidelines for Digital Forensics Laboratories & 2019 & \cite{interpolguidelines19}\\
ENFSI Guidelines  & 2016-2020 &\cite{enfsiguidelines}  \\
     \bottomrule
   \end{tabular}
   }
   \label{tab:mainmodels}
 \end{table}
 
In general, the procedures summarised in Table \ref{tab:mainmodels} have a common hierarchical structure, which can be divided in the steps described in Table \ref{tab:forensicsteps}. Note that, evidence custody changes as well as evidence destruction are two relevant steps that, although not mapped in the high level description provided in Table \ref{tab:forensicsteps}, are included in our system implementation.  

\begin{table}[h]
\rowcolors{2}{gray!25}{white}
\scriptsize
  \centering
   \caption{Main steps in a digital forensic investigation model.}
  \setlength{\tabcolsep}{5pt}
  \begin{tabular}{lp{3.5in}}
    \hline
   \textbf{Forensic Step} &  \textbf{Description} \\
   \hline
Identification  & Assess the purpose and context of the investigation. Initialize and allocate the resources required for the investigation, such as policies, procedures and personnel. \\
Collection \& Acquisition  & The seizure, storage and preservation of digital evidence. Although this two steps need to be strictly differentiated in the physical forensics context, we consider a more relaxed approach in the digital context, since most of times data will be directly collected in a digital form.  \\
Analysis  & The identification of tools and methods to process the evidence and the analysis of the outcomes obtained      \\
Reporting \& Discovery  & The proper presentation of the reports and information obtained during the investigation    to be disclosed or shared with the corresponding entities.   \\
    \bottomrule
  \end{tabular}
  \label{tab:forensicsteps}
\end{table}

\section{Method}

In the following sections, we will describe the case study and the actors involved, our blockchain-based forensic architecture, and the forensic methodology applied to such case. 

\subsection{A case study for an embezzlement scheme.
}

Let Malory be a malicious bank employee who worked at Golden Bank in Albany, Oregon for 15 years. In addition, Malory was promoted to the Head Teller position in the Central Branch five years ago. He was living beyond means; driving a expensive car and being the owner of an expensive home. As he was working for many years in the same branch, he had developed a good relationship with affluent clients of the branch that where living abroad. 

Within such period, the Branch Manager received an oral complaint by the client Alice, who was contesting the balance recorded in the passbook of his saving account. The Branch Manager tried to trail the vouchers of his last transactions but were missing from the physical record. The responsibility of the missing vouchers lied to Malory. The Branch manager, informed the Internal Audit Department of the bank to investigate the case.

The investigation's result was that Malory had embezzled 
500,000 US\$ from a network of 10 clients with similar profile characteristics to Alice (elderly customers living abroad). These clients were served in the branch that Malory was a Head teller. Malory tap into customers' saving accounts to wire funds without authorization for 5 years and proceeded unauthorized withdrawals. Malory used the \textit{cancel transaction} field as a markdown for the embezzled money on each account, so as to keep track what needs to be returned to each account before the client appeared in the branch. 
 
\begin{figure}[th]
    \centering
    \includegraphics[width=\textwidth]{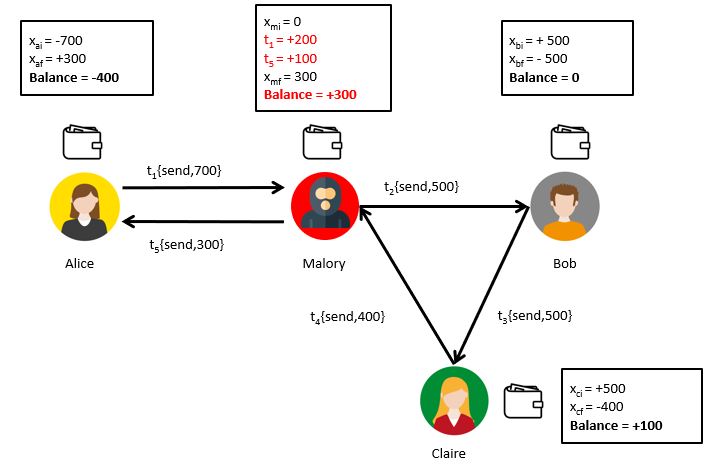}
    \caption{A case study for an embezzlement scheme.}
    \label{fig:cap12}
\end{figure}

Figure \ref{fig:cap12} presents part of Malory's scheme for a selected sample of three clients. In this case, Malory has embezzled a total amount of 300 \$ by performing unauthorized transactions among clients disrespecting the bank's procedures. The next paragraphs describe explicitly three days where Malory acted on behalf of clients and for his own interest.

In Day 1, Malory being alone in the cash desk of the bank's branch performed an unauthorised cash withdrawal of 700 \$ from the banking account of Alice. A minute after he performs an unauthorised cash deposit in the banking account of Bob. From these two transactions he puts 200\$ in his wallet from the cash difference in the bank's ledger. The only evidence from this irregular behaviour is tracked in the CCTV surveillance system, since the physical vouchers linked to unauthorised transactions are lost. At this point of time, Alice has a debit balance of 700 \$ and Bod a credit balance of 500 \$.

Next, in Day 2, Malory performs an unauthorised cash withdrawal of 500\$ from the banking account of Bob, and credits Claire's saving account with an equal amount. At the end of the day, Bob's balance becomes zero while Claire's balance is credited with 500 \$. Similar to Day 1, the only evidence from this irregular behaviour is tracked in the CCTV surveillance system. 

A day later, Malory performs an unauthorised withdrawal from the account of Claire of 400\$ and deposits back to client Alice an amount of 300\$ keeping for himself 100\$. At this point Claire's account has a credit balance of 100\$ and Alice's account has a debit balance of 400\$. Same as above, no voucher bears clients' signatures and the only evidence that can track his irregular behaviour is the  daily CCTV recording.

\subsection{Actors Involved}

\begin{figure}
    \centering
    \includegraphics[trim={0cm 2.4cm 3cm 0cm},clip,width=\textwidth]{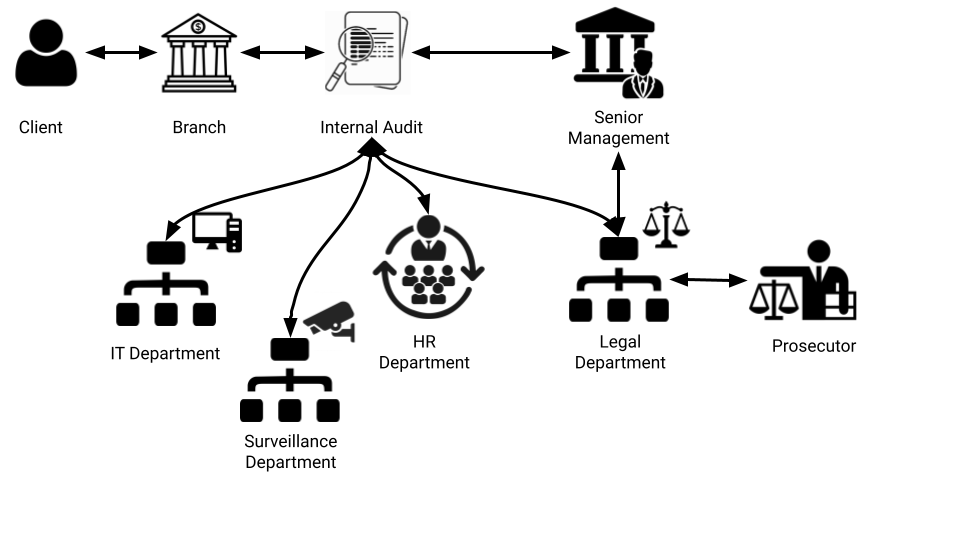}
    \caption{Actors involved in the embezzlement scheme.}
    \label{fig3}
\end{figure}

The main actors involved during the investigation of the above embezzlement scheme are presented in figure \ref{fig3}. Now, we will describe how each actor is involved in the chain of custody.

Firstly, the \textit{branch} archived the written complain of the client contesting the balance recorded in the passbook of his saving account. It was also responsible for the archiving of original vouchers related to the transactions signed by each customer and for the historic record of CCTV material.

The \textit{Internal Audit} received the written declaration of the complaint through the internal mail of the bank and e-mail communication. Original hard copies of the complaint are stored in a separate physical folder linked to Malory's case. The physical folder is scanned at the end of the case and archived in a shared folder with authorized accesses. This folder also includes hard copies linked to the fraud hypothesis (CCTV files, original vouchers, signatures' specimen etc), the signed testimony \footnote{The testimony is recorded as a written agreement between Malory and two investigators (signed by each of them).} of Maroly . The signed report released with investigation results \footnote{A hard copy of the report is distributed internally to Senior management, HR department, and legal department (i.e., these actors are defined later in this section). The recipients sign a mail record which is then archived in the physical folder of the case.}. Digital documentation from detective tools mentioned in \ref{Invest_tech} is archived in the local server of the department, and only authorized users may have access. For the investigation's purposes, the Internal Audit interacts with the bank's IT department, Surveillance Department, the Branch, and Human Resources (HR) Department. 

\textit{IT Department} assists Internal Audit by providing data from the database of core banking system.
The data are sent via e-mail usually as an excel file concerning transactions and daily journal of entries of Malory. Also, when Internal Audit needs to trace additional CCTV files, it collaborates with \textit{Surveillance Department}. The latter delivers any CCTV files upon request through USB. Each employee has a physical record with personal information archived in \textit{HR Department}. This information distributed to investigators through emails. Also, it is responsible for the archiving of disciplinary board's proceedings. 

When the investigation's report is issued, \textit{Senior Management} receives it. Its role is to decide and approve any disciplinary  and litigation actions. Decisions and approvals are documented and archived. If the decision involves litigation actions then \textit{Legal Department} undergoes the submission to a prosecuting authority. 

At last, the \textit{Prosecutor} receives a structured package including the disciplinary board's decision, the findings' report and a copy of physical evidence and proceeds to further legal actions.

\subsection{Forensic Procedure}
\label{sec:forensic}

In the case of chain of custody and trail of events preservation, we need to ensure that our system enables features such as integrity, traceability, authentication, verifiability and security \citep{Bonomi2020,Tian2019}. In this regard, Table \ref{tab:chainofcustody} provides a description of each feature and how our blockchain-based system enables it. 

\begin{table}[h]
\rowcolors{2}{gray!25}{white}
\scriptsize
  \centering
   \caption{Main features required to guarantee chain of custody preservation.}
  \setlength{\tabcolsep}{5pt}
  \begin{tabular}{lp{3.5in}}
    \hline
   \textbf{Feature} &  \textbf{Description} \\
   \hline
Integrity  & The events data as well as evidences cannot be altered or corrupted during the transferring and during analysis due to the use of hashes. \\
Traceability   & The events and evidences can be traced from their creation till their destruction since every interaction is stored in an immutable ledger.    \\
Authentication  & All the actors and entities are unique and provide an irrefutable proof of identity due to the use of asymmetric cryptography.   \\
Non-repudiation & Each action can be related with its author, enabling strong accountability guarantees. \\
Verifiability  & The transactions and interactions can be verified by the corresponding actors.
This verification can be performed in real time.     \\
Security & Only actors with clearance can add content or access to it. A robust underlying consensus mechanism ensures that the transactions are signed in a cryptographically secure way. \\
    \bottomrule
  \end{tabular}
  \label{tab:chainofcustody}
\end{table}

\begin{figure}[th]
    \centering
    \includegraphics[trim={0cm 3cm 0cm 0cm},clip,width=\textwidth]{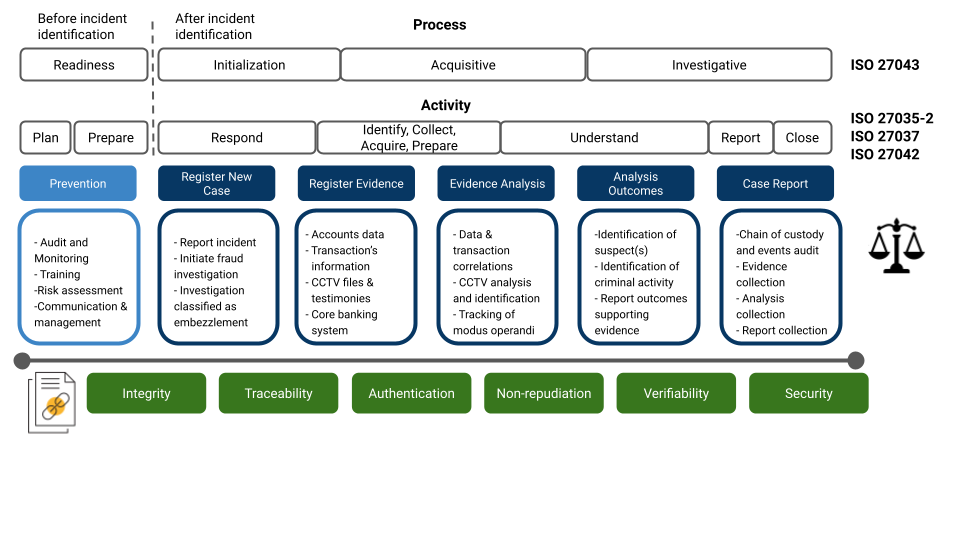}
    \caption{Embezzlement Forensic investigation main phases according to standardised procedures (top) and the corresponding actions performed in our case study (bottom).}
    \label{fig:flowflorensics}
\end{figure}

In addition, Figure \ref{fig:flowflorensics} summarises the main tasks performed in each investigation phase according to our case scenario, and their corresponding relationship with the forensic flow. We included the process defined in ISO 27043:2015 \citep{iso270432015}, as well as each step defined in the guidelines to plan and prepare for incident response (ISO/IEC 27035-2:2016 \citep{iso27035}), the guidelines for the identification, collection, acquisition and preservation of digital evidence (ISO/IEC 27037:2012 \citep{iso27037}) and the guidelines for interpretation and analysis of digital evidence (ISO/IEC 27042:2015 \citep{iso27042}). Therefore, we mapped the different steps of the investigation as defined in our method. Note that we included the prevention layer in our design, which details will be later discussed in Section \ref{sec:discussion}. Therefore, after reporting the incident and initiating the investigation, the evidence collection and forensic analysis is summarised in following steps:

\begin{enumerate}
    \item Collection and analysis of the investigated accounts (including saving accounts of clients and their correlation with employees' accounts).
    \item Analysis of the daily transactions recorded in the journal of entries of Malory (i.e. extractions from core banking system).
    \item Reconcile the time of transactions appearing in the journal entry with the CCTV time.
    \item Review of CCTV files in order to trace the physical presence of the client and the suspect.
    \item Collection of the testimony of the suspect(s) in signed hardcopy.
\end{enumerate}

For an investigation to be sound, all the forensic steps need to be provable and, in the case of evidence analysis, results need to be reproducible. In the case of internal audits, a malicious investigator could tamper evidence or the analysis performed on data to hide proofs. Therefore, a robust forensic procedure is required to guarantee that evidence is collected in a sound manner and it is not tampered during the analysis. Moreover, each forensic action has to be paired with an individual. The aforementioned requirements can be accomplished by means of the blockchain-based forensic architecture described in the next section.

\subsection{Blockchain-based Forensic Architecture}

In this section, we describe our blockchain-based forensic architecture. In our setup, we assume that the system is implemented in the context of a secure laboratory/investigation facility according to a set of policies and regulations. Note that, since each region and country may apply different policies, we leave their definition and discussion as a future research line. Nevertheless, our system can accommodate more functionalities in the smart contract definition, as well as higher layer control systems and application programming interfaces (API)s. 

\begin{figure}[th]
    \centering
    \includegraphics[trim={0cm 5.4cm 0cm 0cm},clip,width=\textwidth]{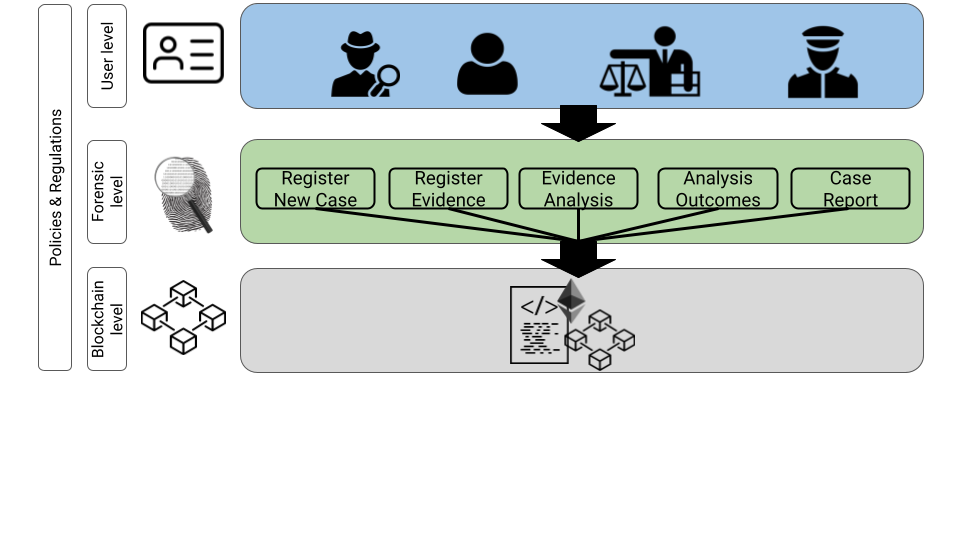}
    \caption{Overview of the different levels of the system.}
    \label{fig:architecture}
\end{figure}

Considering the previously stated forensic flows and the characteristics of the embezzlement scenario, the architecture of our method is depicted in Figure \ref{fig:architecture}. Different policies and regulations will be applied at each level identified in Figure 5, so that, e.g. the identity management, including roles and permissions, will be defined by the corresponding authorities according to each jurisdiction. Moreover, the forensic guidelines and standards, as well as the underlying blockchain technology to be used and the definition of the smart contracts are also tied to the same principle.

The first step involves the case creation. In this regard, a case can be registered due to a citizen's testimony (we include in this definition any individual that wants to report a crime) or directly by a prosecutor (or an investigator with enough clearance to open a case) who observed suspicious behaviour. Next, evidences are collected and analysed by using the appropriate forensic tools. The description of each action (e.g. storing an evidence and analysing an evidence) may have a description file associated with JSON or CSV format, to ease further searches and classifications. We assume that secure and private storage is used to preserve the evidences, but other platforms such as cloud-based storage or decentralized storage systems such as the Inter-Planetary File System (IPFS) \citep{benet2014ipfs} can be used if data are properly protected/encrypted (i.e. following the definitions set out in information security standards like ISO/IEC 27001 \citep{ganji2019approaches} or other national IT-security guidelines). More concretely, depending on the approach selected by the investigators, the hashes of the evidence can point to an IPFS address or record the SHA-256 hash of the evidence, the latter being the prevalent method in most threat intelligence platforms, such as Virustotal\footnote{https://www.virustotal.com/} or MalwareBazaar\footnote{https://bazaar.abuse.ch/}. Therefore, our proposed system enables the extraction of the hashes of an investigation via a set of smart contract functions, as later described in Section \ref{sec:experiments}, enabling investigators to use such intelligence services (e.g. by using their APIs to upload the evidence or by a hash query) to retrieve additional intelligence. Despite its practicality, an automated methodology to integrate a query system for each intelligence platform requires different configurations and is left to future work.  

Finally, when the investigation concludes, all the data can be collected and presented in court.

The aforementioned interactions are mapped into a smart contract and therefore stored permanently in the blockchain. The latter guarantees the verifiability of the investigation due to the blockchain's immutability, as well as the preservation of the chain of custody, as mentioned early in Table \ref{tab:chainofcustody}. Therefore, the investigation can be audited to certify that any evidence was tampered during the investigation, guaranteeing the soundness of the different forensic procedures. In addition, our approach is designed to be accommodated and in other digital investigation contexts apart from embezzlement, enhancing its adaptability to internal audits.

Concerning the identity management scheme, we argue that due to each organisation's specific regulatory requirements and policies, which may entail further definitions, agreements and developments, the definition of such scheme falls out of the scope of this paper. Therefore, we consider the identity management module as a black box in our architecture to apply standard and validated mechanisms. For instance, further than the MetaMask \footnote{http://metamask.io/} web3 plugin used in our testing setup to manage the wallets and operating the smart contracts through our web interface, other blockchain-based identity management systems could be applied in this layer. Blockchain circumvents the boundary-based digital identity problem by delivering a secure solution without the need for a trusted, central authority managing access permissions through smart contracts. Since blockchain is considered one of the main enablers of self-sovereign identities, alongside with verifiable credentials and decentralised identifiers, there are multiple examples of privacy-preserving blockchain-based identity management systems and already functional projects \citep{jacobovitz2016blockchain,8425607,s18124215} that could be adopted for our forensic platform. In this regard, some approaches enabling multi-authority attribute-based access control with smart contracts have been presented in the literature \citep{guo2019multi,guo2019flexible}, as well as approaches implementing multi-blockchain approaches for fine-grained access control \citep{9146294}.
In addition, FIDO-based authentication mechanisms could also be used to enable higher security standards \citep{morii2017research,lyastani2020fido2}, including biometric access control, criptographically secure credentials, and passwordless authentication.

\section{Experiments and Integration Details}
\label{sec:experiments}
\begin{figure}[th]
    \centering
    \includegraphics[width=\columnwidth]{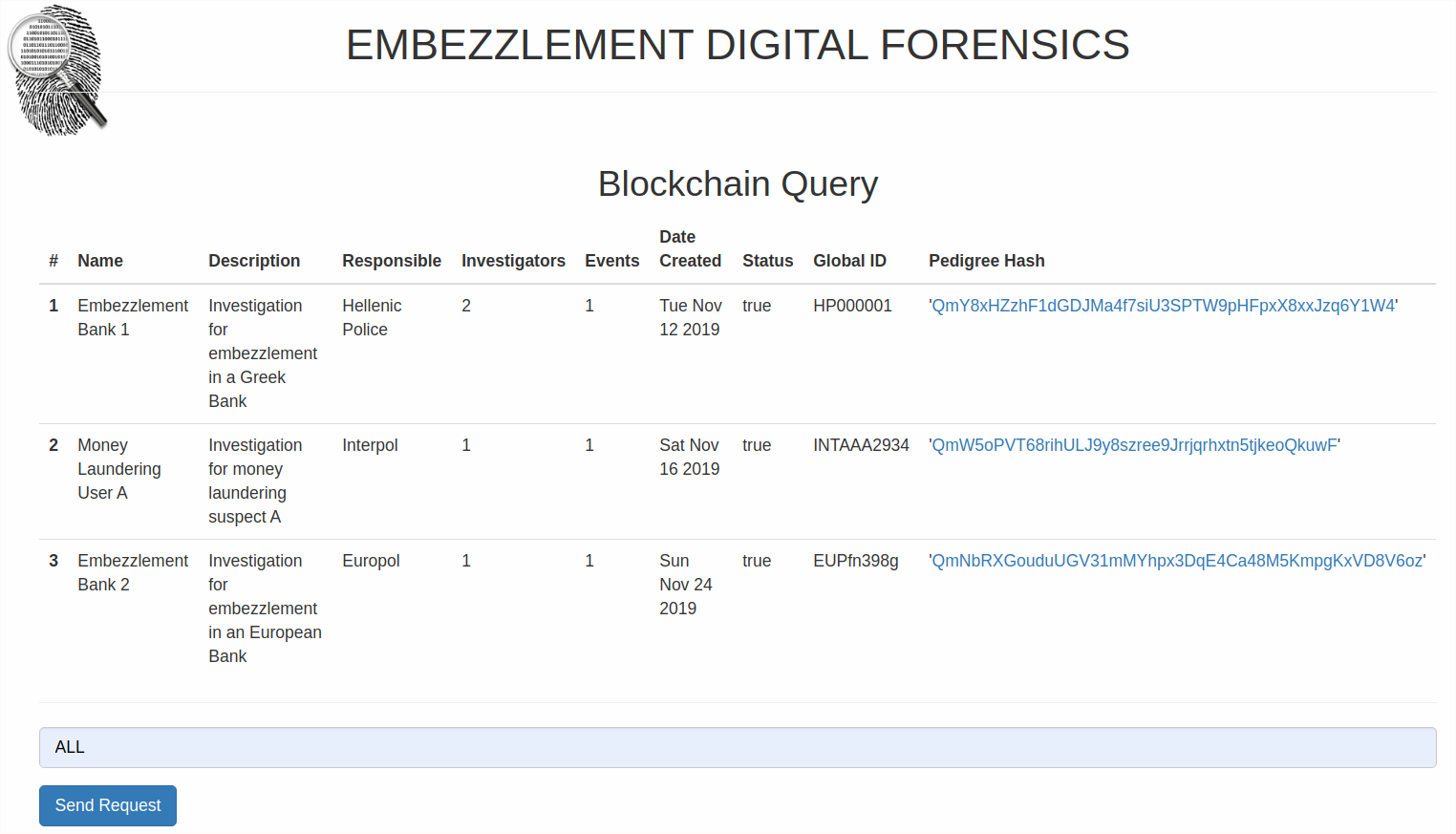}
    \caption{Example of the outcome of a query by searching all cases and their corresponding IPFS links.}
    \label{fig:graphical}
\end{figure}

\begin{figure}[th]
    \centering
    \includegraphics[width=\columnwidth]{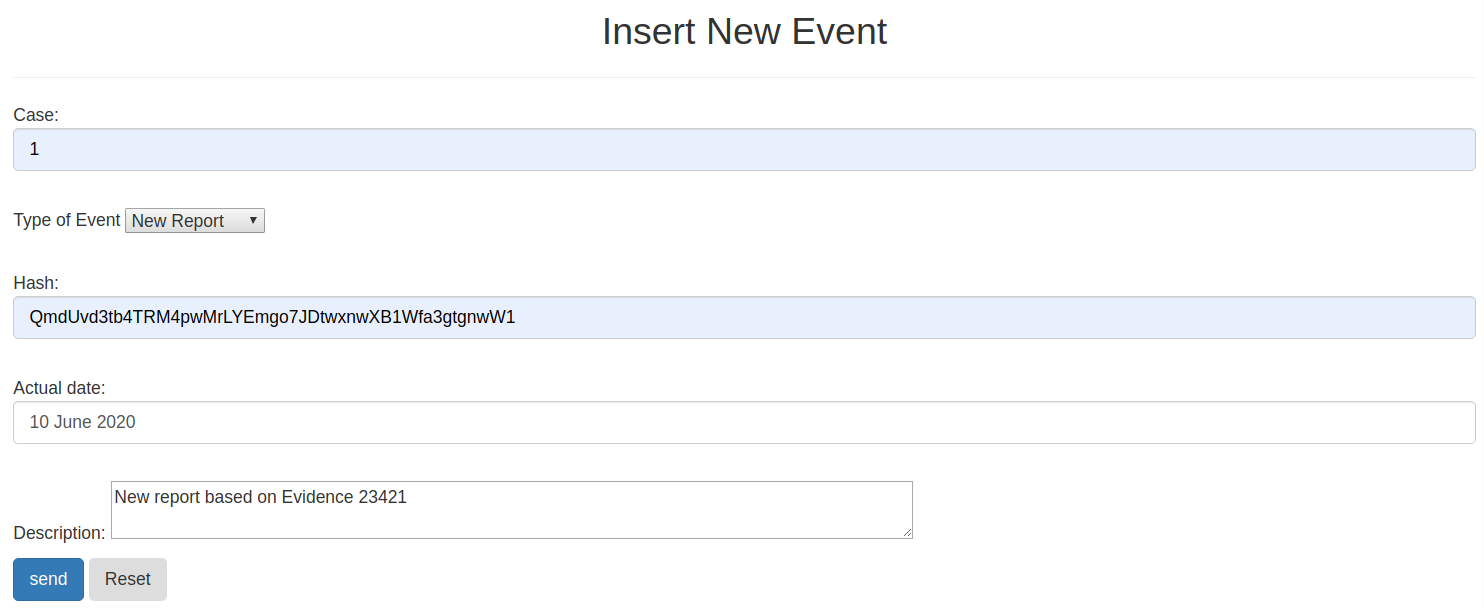}
    \caption{A detail of the form user to insert a new event to the system.}
    \label{fig:graphical2}
\end{figure}

The interactions between the different actors of the system and the forensic events have been implemented by means of a smart contract, and different tests have been performed in a local private blockchain to showcase the feasibility and performance of the proposed method. More specifically, an Ethereum-based blockchain using \texttt{node}\footnote{https://nodejs.org/} and \texttt{ganache-cli}\footnote{https://github.com/trufflesuite/ganache-cli} was created, and  \texttt{truffle}\footnote{http://truffleframework.com} was used to code and deploy a fully functional smart contract. Moreover, a graphical interface was developed in order to query and insert information stored in the blockchain by using node package manager \texttt{npm} \footnote{https://nodejs.org/en/knowledge/getting-started/npm/what-is-npm/}, which also retrieves the corresponding hash of the directory of a specific investigation and its link to the IPFS \citep{benet2014ipfs}, along with other detailed information. Therefore, the information of a specific investigation (or a set of them) is graphically depicted for the user, as well as the option to store a new event, as seen in Figures \ref{fig:graphical} and \ref{fig:graphical2}, respectively.

We have selected Ethereum due to its robust consensus mechanism as well as its smart contract framework (i.e. we use solidity 0.5.0, which prevents vulnerabilities such as \textit{uninitialized storage pointer} and \textit{erroneous visibility}). For more information about the security of Ethereum and its smart contract framework, we refer the interested reader to \cite{xiao2020survey} and \cite{chen2019survey}, respectively.

For the sake of completeness, a detailed description of the functions implemented in the smart contract is provided in Table \ref{tab:functions}. As an additional feature, every time the information of the smart contract is updated, a \textit{trigger} function is called, which can be used as an alert. Therefore, the investigators will be able to check in real time the information about a given investigation using the \textit{get/retrieve} functions implemented in the smart contract for verification or managerial purposes.

\begin{table*}[h]
\rowcolors{2}{gray!25}{white}
  \centering
   \caption{Main characteristics and permissions of the functions implemented in the smart contract.The permissions column states which functions are public (P) or restricted (R) only for specific users related to a case.}
  \setlength{\tabcolsep}{5pt}
  \resizebox{\textwidth}{!}{
  \begin{tabular}{lcp{2in}ccp{2in}}
    \hline
   \textbf{Code} &  \textbf{Function} & \textbf{Input} & \textbf{Output} & \textbf{Permissions} & \multicolumn{1}{c}{\textbf{Description}} \\
   \hline
$f1$  & constructor                    & void                                                                  & na                   & na          & creates the smart contract                           \\
$f2$  & addCase                     & String name, String description, String responsible, String globalID, timestamp, hash & void                 & R         & adds a new case to the system                   \\
$f3$ &    updateCaseDescription  &   Uint caseID, string description    &      void  &  R   & updates a case description  \\
$f4$ &    updateCaseStatus  &   Uint caseID, string status    &      void  &  R   & updates a case status  \\
$f5$ &    updateResponsible  &   Uint caseID, address responsible    &      void  &  R   & updates a the responsible of a case  \\
$f6$ &    addInvestigatorCase  &   Uint caseID, address investigator    &      void  &  R   & adds a new investigator to a case\\
$f7$ &    getNumberOfCases  &   void   &      Uint  &  P   & returns the global number of cases \\
$f8$ &    getCase  &   Uint caseID    &      Object  &  P   & returns a case object and its information  \\
$f9$ &    getCaseGlobalID  &   Uint caseID   &      String  &  P   & returns the global ID of a case \\
$f10$ &    getNumberofInvestigators  &   Uint caseID    &      Uint  &  P   & returns the number of investigators asigned to a case  \\
$f11$ &    getCaseHash  &   Uint caseID    &      hash  &  P   & returns the hash pointer with information of a case  \\
$f12$ &    addEvent  &   Uint caseID, string type, String description, String status, hash, timestamp    &      void  &  R   & adds a new event to a case  \\
$f13$ &    updateEventStatus &   Uint eventID, string status    &      void  &  R   & updates an event/evidence status  \\
$f14$ &    getNumberOfEventsCase  &   Uint caseID   &      Uint  &  P   & returns the number of events of a case  \\
$f15$ &    getEventsCase  &   Uint caseID    &      Object &  P   & returns the set of events related to a case  \\
$f16$ &    getGlobalNumberOfEvents  &   void    &      Unit &  P   & returns global number of registered events  \\
$f17$ &    getEvent  &   Uint eventID    &      Object &  P   & returns an event object  \\
$f18$ &    getEventHash  &   Uint eventID    &      Hash &  P   & returns the hash of a specific event \\
$f19$ &    triggers  &   void    &      void &  P   & trigger functions to update the status of the smart contract  \\
    \bottomrule
  
   \end{tabular}
  }
  \label{tab:functions}
\end{table*}

A link between the smart contract's functions and the forensic procedures is summarised in Table \ref{tab:codeprocedures}. In the case of custody change procedures, the involved actors will be stored in the corresponding list of owners of the evidence along with a timestamp. Moreover, a new event can be created with an evidence pointing to the same hash, yet with a new creation date to grasp the import of such evidence. In the case of the destruction of an evidence, this procedure will change the status of the evidence to \textit{deleted}. In the latter case, the deletion of a file in a local storage is an easy task, yet the complete erasure of the content in e.g. IPFS can only be made feasible through novel mechanisms \citep{POLITOU2020956}, since IPFS does not implement an erasure protocol at the time of writing \citep{ipfslastrelease,casino2019immutability}. In the case of blockchain, the problem is exacerbated due to its inherent immutability \citep{politou2018forgetting,politoublock,deuber2019redactable}, and the best solution is to minimise the direct storage of data (i.e. only pointers and hashes).

\begin{table}[ht]
   \setlength{\tabcolsep}{7pt}
   \scriptsize
   \caption{Relationship between smart contract functions and forensic procedures.}
  \begin{tabular}{cc}
     \toprule
   \textbf{Forensic embezzlement procedure} &  \textbf{Code}  \\
   \midrule
    Register new case           &    $f1,f2,f3$     \\
    Evidence management          &   $f12$, $f13$   \\
     Evidence analysis          &    $f8$, $f11$, $f12$, $f17$, $f18$    \\
     Analysis \& Report           &  $f8$, $f9$,  $f10$,  $f11$, $f14$, $f15$, $f18$   \\
     Admin \& Statistics &   $f4$, $f13$, $f5$, $f6$, $f7$, $f9$, $f10$, $f14$, $f15$, $f14$, $f17$, $f19$  \\
     \bottomrule
   \end{tabular}
   \label{tab:codeprocedures}
 \end{table}

In addition to the benefits enabled by the smart contracts, it is essential to guarantee the privacy of the transactions and the involved actors. Therefore, the contents can be modified only by participants with specific roles (each function is implemented with concrete permissions, e.g. using the \textit{require} clause of solidity and variables such as \textit{msg:sender} to check account authenticity), thus enabling secure access control. For example, prosecutors will be able to open and close investigations, but investigators will be able only to add new events to a specific investigation. In our setup, read-only functions and variables can be checked by public users. Nevertheless, more sophisticated access control, policies and data protection measures can be implemented by the users of the platform according to their specific needs. 

The transactions tested in the developed private blockchain (e.g. deployment of the smart contract, adding a new case and adding a new event) are performed in the order of milliseconds, and thus, our approach enables real-time interactions.
The implementations as well as the graphical interface web service are available on GitHub\footnote{\url{https://github.com/francasino/financial_forensics}}. 

\section{Discussion}
\label{sec:discussion}
In this section, we discuss both the benefits and the challenges to be overcome in the blockchain digital forensics field. Moreover, we provide a granular analysis of such challenges across the different topics explored in this work, namely, digital forensics, blockchain, and finance related crimes.

\begin{table}[ht]
   \setlength{\tabcolsep}{7pt}
   \scriptsize
   \caption{Prevention mechanisms and their corresponding application level.}
   \resizebox{\textwidth}{!}{
  \begin{tabular}{p{.75\textwidth}l}
     \toprule
   \textbf{Mechanism}  &  \textbf{Level affected} \\
   \midrule
Training on anti-fraud and appropriate behavioural conduct & Workers \\
Efficient networking and communication strategies & All \\
Transparent managerial practices & Executives \\
Risk assessment to both internal and external fraud & All \\
Integration of security and AI experts in the organization charts & Workers \\
Audit controls and appropriate resources & All \\
     \bottomrule
   \end{tabular}
   }
   \label{tab:prevention}
 \end{table}
 
 According to the challenges extracted from our literature review, we identified several mechanisms that could be used to prevent/minimise financial fraud in institutions and their corresponding context of application in Table \ref{tab:prevention}. As it can be observed, training on anti-fraud and incorporating higher security standards are necessary to help employees in their daily activities. In addition, executive personnel should implement more comprehensive and transparent managerial practices to avoid the possibility of obscure and unethical activities. Finally, communication, networking and risk assessment are crucial to improving the information flows between all the actors involved.

Blockchain and its benefits to a myriad of application scenarios have been thoroughly discussed in the past \citep{casino2019systematic,kuo2017blockchain,OLNES2017355,zheng2018blockchain}. Features such as immutability, verifiability, auditability, security were enhanced by the automation provided by smart contracts. For example, the existence of an event at an specific time (e.g. the existence of a file, a good, a token or any kind of asset) can be verified in a matter of milliseconds, due to the self-executing capabilities and the real time synchronization of the information in the blockchain. The latter enables proof-of-existence when paired with the use of hashes, which can be used to guarantee the trail of events as well as the proper preservation of the chain of custody in the digital forensics context.

In addition, the benefits go further beyond the provision of a solid proof in court, since the knowledge an evidence gathered in a case can be correlated in the future to reduce the time required to find a vulnerability or speed up cybercrime investigations. This is particularly relevant in the finance context, where all background and identification information can be stored. For instance, the Know your Customer (KYC) can be processed easier and faster during investigations and be secured against any internal fraudulent activities. On top of that, the application of smart contracts could prevent efforts for forgery and counterfeit documents. Therefore, the use of blockchain and its data mining capabilities will foster collaboration between different entities to share information and enable the early detection of embezzlement schemes. Moreover, evidences and reports can provide valuable input for the elaboration of AI models with which increase the rate of embezzlement detection, since embezzlement schemes sometimes last for more than 5 years.

 As discussed in \citep{CHANG2020120166}, the authors identified several critical success factors for adopting blockchain technology in the finance sector. In general, robust and efficient blockchain implementations are critical to guarantee the required infrastructure to leverage financial services. The latter, paired with well-trained teams and robust security and privacy guarantees, will close the gap for the adoption of blockchain in finance. Other crucial aspects are reputation and community building and mechanisms to guarantee upgraded, long-term products and services. 
Another relevant aspect that hinders blockchain adoption is the lack of global regulations \citep{ALI2020102199}, due to highly regulated industries paired with complex jurisdictional and legal frameworks. In addition, underdeveloped countries with insufficient technological infrastructure, and an unclear privacy management and data governance are also delaying blockchain's adoption \citep{kim2017does}. The latter is translated into a lack of interoperable solutions and reactive behaviours displayed by regulators, which are always a step behind novel trends and opportunities. 

 The proper use and adoption of blockchain and showcasing its potential is vital. We believe that systems like the one proposed in this paper reinforce the trust in blockchain and close the gap between the financial sector and blockchain, not only in the context of digital investigation but as a potential tool to leverage better services backed with solid regulatory frameworks supporting them.

Despite the aforementioned benefits, blockchain is not a panacea. There are still many challenges to overcome, such as the issues related with standardization of the procedures related with digital crime prosecution. The latter includes cross border investigations, data storage and data sharing policies, and the management of personal data. Note that the storage of personal data in a blockchain contradicts the GDPR and the right to be forgotten \citep{politou2018forgetting,politoublock} due to immutability. Therefore, despite the unarguable benefits of the immutability property, the capability of sharing an asset perpetually can be used with malicious ends as already seen in the literature \citep{li2020survey,conti2018survey,ariadnedns}. Moreover, this issue is not only affecting blockchain but also decentralised permanent storage solutions, due to the lack of effective erasure mechanisms \citep{casino2019immutability}. We believe the controlled erasure of data in blockchain, as well as similar systems, will be a relevant research area in the near future and some authors are already investigating it \citep{politoublock,radinger2020blockchain,deuber2019redactable,9060994}.

Another relevant flaw of blockchain is its scalability, which varies according to the volume of transactions, verifiers/miners, consensus mechanism and other features. In this regard, it is well-known that private flavours of blockchain are the most suitable for most of applications, yet their scalability when the number of users and interactions grow also suffers performance issues \citep{dinh2017blockbench,pongnumkul2017performance,dong2019dagbench,POLITOU2020956}. 

In our implementation, we minimise the amount of data stored in the blockchain by using hashes. Nevertheless, our testing protocol focused on the underlying functionalities in order to showcase the feasibility of robust digital investigation and verifiable chain of custody. Therefore, the implementation of an advanced identity management system and the required policy definitions to overcome the issues of cross border investigations are not covered in this paper, and are left for future research.

The key benefit from the managerial perspective of our proposed architecture is that it can eliminate operational risk and costs associated with investigation procedures. In particular, it can accelerate investigations by facilitating the reconciliation of evidence in a verifiable and auditable manner, increasing the possibility of identifying perpetrators and reducing fraud management and recovery costs. It enables senior management to have direct access to fraud risk exposure, reporting, and monitoring and gain a comprehensive view of all investigative information improving their decision-making analysis to impose sanctions and proceed to litigation actions. Moreover, the proposed architecture can be adopted for the audit trail of regularly planned internal audits.

Based on the same methodological steps, the auditors can store their evidence from their audit trails to better monitor their follow-up actions. Thus, regular internal audits can rip the benefits of this architecture and deliver efficient outcomes to the management. Blockchain adoption might have a double effect on the internal audit department since it can reduce operational costs associated with the streamlining of auditing and investigative procedures.

\section{Conclusions}

In this paper, we recall the negative impact of financial crimes on society, focusing on embezzlement scenarios. We provide a thorough state-of-the-art of blockchain-based digital forensic approaches, increasing their relevance in such investigations due to their inherent features. Next, we propose a functional implementation of a forensically sound flow to investigate financial crimes based on Ethereum, and we test it with a real-world embezzlement use case. The outcomes showed that our proposal empowers integrity verification, tamper-proofness, and adaptability towards other fraud and financial crimes. 

Despite the benefits of our approach, we leveraged a profound analysis of the literature, identifying the fundamental challenges for adopting blockchain and the financial sector and proposing some strategies to overcome them. Moreover, we discussed further benefits of our approach and how it enhances some of the features required for closing the gap between blockchain and the financial sector.

Among the most important aspects of blockchain integration in financial investigations - and internal audits - are the guarantee of invariable evidence and the streamlining of auditing and investigation procedures. The management of a financial institution may reduce several operational costs associated with the fraud incident and its investigation. The whole framework facilitates the interaction of interdisciplinary resources securely and confidentially and provides a comprehensive overview of each organization's management related to monitoring and reporting fraud risk exposure.
Future work will focus on increasing the adaptability of the scheme by enabling other functionalities and the migration to other blockchain systems. Moreover, we aim to provide a robust identification mechanism to enable secure and private cross-border investigations.

\begin{acknowledgements}
This work was supported by the European Commission under the Horizon 2020 Programme (H2020), as part of the project \textit{LOCARD} (\url{https://locard.eu}) (Grant Agreement no. 832735) and \textit{YAKSHA}, (\url{https://project-yaksha.eu/project/}) (Grant Agreement no 780498).
\end{acknowledgements}

\bibliographystyle{spbasic}
	
	\bibliography{Bibliography.bib}

\end{document}